V. Bakhrushin

# SOFTWARE REALIZATION OF THE COMPLEX SPECTRA ANALYSIS ALGORITHM IN R

Abstract. Software realization of complex spectra analysis algorithm is proposed, which is based on using the non-linear minimization of the sum of squared residuals and set of criteria, which check their statistical properties.

**Formulation of the problem**. The problem of complex spectrum decomposition on the elementary components is actual for a wide range of tasks in materials research, technical diagnostics etc. In particular, it arises in studying of mechanical relaxation [1], photoluminescence [2], deep level transient spectroscopy in semiconductors [3] and the like. A characteristic feature of this problem is that each component has a defined physical meaning. In particular, they may correspond to different processes in crystals, different states of impurities and defects and so on. Therefore, an adequate model of the spectrum should not contain redundant components. In this regard, an important task is not only correct approximation of the spectrum shape, but also the correct determination of the number of components and their parameters.

**Analysis of recent research and publications**. There are different approaches to solving the problem of the complex spectrum decomposition into components [4, 5]. In this case, one of the major difficulties is the need to choice the correct number of components. In [1, 6] for its solution it has been proposed to use a system of adequacy criteria that evaluate conformity of model residuals with normal distribution, equality to zero of their mean value and their statistical independence. In this case, it may be proposed loop which supposes the following main steps:

− specifying the number of components (it can be taken equal to one);
− specifying the initial values of components' parameters;
− refinement of the components' parameters using the methods of nonlinear minimizing the sum of squared residuals of the model;
− check of model adequacy using a set of criteria;
− if a model is inadequate - increasing the number of components by unity and repetition of the subsequent procedure.

**Statement of an objective**. To develop a software implementation of this algorithm using R.

**Basic material of research.** R language is widely used in the field of applied statistics and data analysis [7]. It is caused by the large number of specialized functions and libraries designed to solving such problems, as well as by the fact that R is a freeware product with open source. This allows unlimited development of programs for the implementation of new algorithms and methods.

Model of the complex spectrum representing the sum of n Debye peaks, can be written [1] as:

$$Q(T) = \sum_{j=1}^{n} Q0_j \, cosh^{-1}\left[\frac{E_j}{R}\left(\frac{1}{T} - \frac{1}{T0_j}\right)\right], \quad (1)$$

$$E_j = RT0_j \ln\frac{k_b T0_j}{hf}, \quad (2)$$

where $Q0_j, E_j, T0_j$ – components parameters, $R, k_b, h, f$ – constants.

For the forming of the spectrum such function was written:

```
Q = function(T, Q0, T0) {
E = R*T0*log(kb*T0/h/f)
QQ = matrix(nrow = length(T), ncol = length(T0))
for (j in1:length(T0))
QQ[,j] = Q0[j]/(cosh(E[j]/R*(1/T - 1/T0[j])))
rowSums(QQ)
}
Q(T, Q0, T0)
```

It uses numerical vectors $T, Q0, T0$ and returns the values of numerical vector $Q$, which correspond to values of $T$.

Model is built using function nls(). For it correct work we must specify a data array Qemp and initial values of parameters:

```
Data = data.frame(T, Qemp)
names(Data) = c("T", "IFr")
mod = nls(IFr~Q(T, Q1, T01), Data, start = list(Q1 = c(1, 1, 1), T01 = c(450, 550, 650)))
```

Function nls() in this case evaluates the model parameters by minimizing the sum of its squared residuals using Newton algorithm. However, it may be specified arguments that establish other methods of evaluation.

The next step is to evaluate the adequacy of the model. To verify the residuals compliance with normal distribution it is used the Anderson – Darling criterion implemented with the function ad.test() from the library "nortest". The equality of residuals arithmetic mean to zero is checked by One-Sample t-test using the function t.test(). For verification of autocorrelation of the first five orders the function durbinWatsonTest() from library "car" is used:

```
library(nortest)
ad.test(resid
t.test(resid)
library("car")
durbinWatsonTest(lm(resid~T), max.lag = 5)
```

The difference of variances of model residuals and errors of empirical data also is an important indicator of its inadequacy. It may be checked by Fisher test. But the standard function var.test() is not applicable in this case, since it ignores the number of spectrum model parameters. Therefore, for this purpose it can be used an alternative approach:

```
var_res = sum(((resid - mean(resid))^2)/(length(resid)-length(Q1)))
FF = max(var_eps, var_res)/min(var_eps, var_res)
p_value = 1-pf(FF, length(T) - length(Q1), length(T) - length(Q1))
```

Examples of spectrum decomposition for the cases of right and wrong choice of the number of components are shown in Fig. 1, 2.

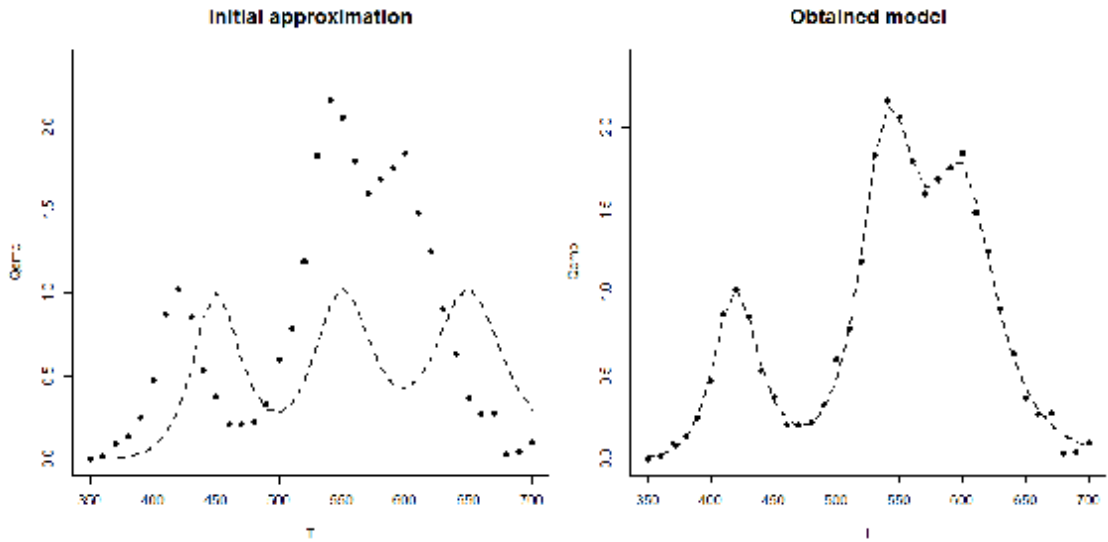

Fig. 1 – The results of the spectrum decomposition with the right choice of the number of components

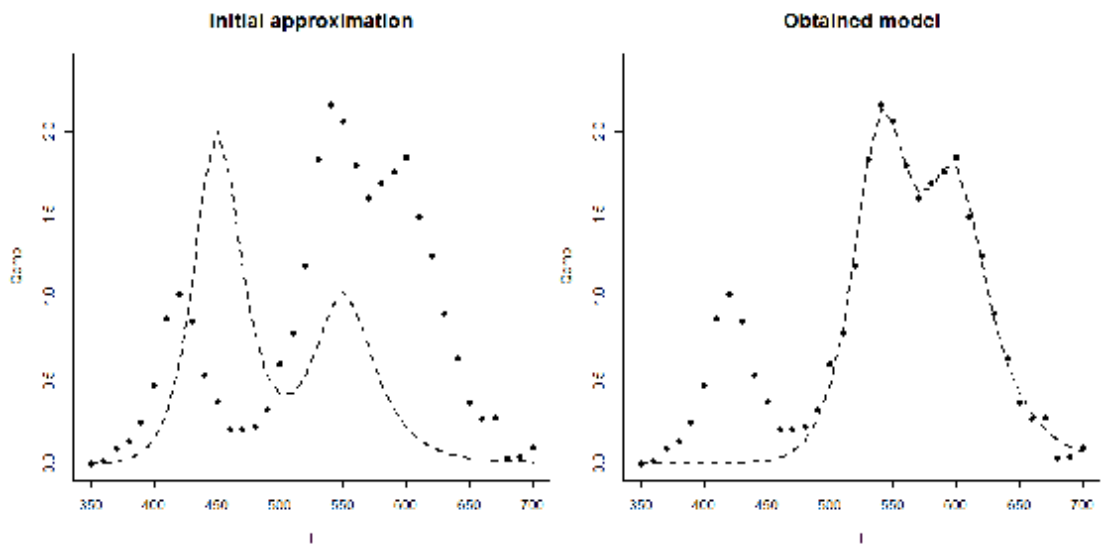

Fig. 2 – The results of the spectrum decomposition with the wrong choice of the number of components

Conclusions. Software implementation of the algorithm of the complex spectrum decomposition on unknown number of Debye components is proposed. Testing has shown the correctness of the program for a wide range of conditions, corresponding practically important situations.

# REFERENCIES